\documentclass[iop]{emulateapj}

\usepackage{natbib}

\shorttitle{Sensor Distortion Effects}
\shortauthors{Peterson et al.}

\begin{document}

\title{Sensor Distortion Effects in Photon Monte Carlo Simulations}

\author{J.~R.~Peterson$^1$, P.~O'Connor$^2$, A.~Nomerotski$^2$, E.~Magnier$^3$, J.~G.~Jernigan$^4$, J.~Cheng$^1$, W.~Cui$^{1,5}$, E.~Peng$^1$, A.~Rasmussen$^6$, G.~Sembroski$^1$}

\affil{$^1$~Department of Physics and Astronomy, Purdue University, West Lafayette, IN 47907, USA}
\affil{$^2$~Brookhaven National Laboratory, Upton, NY 11973, USA}
\affil{$^3$~Department of Astronomy, University of Hawaii, Honolulu, HI 96822, USA}
\affil{$^4$~Eureka Scientific, Oakland, CA 94602, USA}
\affil{$^5$~Tsinghua Center for Astrophysics, Department of Astronomy Tsinghua University, Beijing 100084, China}
\affil{$^6$~Stanford Linear Accelerator Laboratory, Menlo Park, CA 94025, USA}

%\author{J.~R.~Peterson}
%\affil{Department of Physics and Astronomy, Purdue University, West Lafayette, IN 47907, USA}

%\author{P.~O'Connor}
%\affil{Brookhaven National Laboratory, Upton, NY 11973, USA}

%\author{A.~Nomerotski}
%\affil{Brookhaven National Laboratory, Upton, NY 11973, USA}

%\author{E.~Magnier}
%\affil{Department of Astronomy, University of Hawaii, Honolulu, HI 96822, USA}

%\author{J. G. Jernigan}
%\affil{Eureka Scientific, Oakland, CA 94602, USA}

%\author{J. Cheng}
%\affil{Department of Physics and Astronomy, Purdue University, West Lafayette, IN 47907, USA}

%\author{W. Cui}
%\affil{Department of Physics and Astronomy, Purdue University, West Lafayette, IN 47907, USA}
%\affil{Tsinghua Center for Astrophysics, Department of Astronomy Tsinghua University, Beijing 100084, China}

%\author{E.~Peng}
%\affil{Department of Physics and Astronomy, Purdue University, West Lafayette, IN 47907, USA}

%\author{A.~Rasmussen}
%\affil{Stanford Linear Accelerator Laboratory, Menlo Park, CA 94025, USA}

%\author{G.~Sembroski}
%\affil{Department of Physics and Astronomy, Purdue University, West Lafayette, IN 47907, USA}

%\correspondingauthor{John~R. ~Peterson}
\email{peters11@purdue.edu}

\begin{abstract}

  We present a detailed method to simulating sensor distortions using a photon and electron Monte Carlo method.  We use three dimensional electrostatic simulations to parameterize the perturbed electric field profile for non-ideal sensor details.  We follow the conversion of simulated photons, and the subsequent response of the converted electrons to the electric field pattern.  These non-ideal sensor details can be implemented efficiently in a Monte Carlo approach.  We demonstrate that the non-ideal sensor distortions have a variety of observable consequence including the modification of the astrometric pattern, the distortion of the electron diffusion size and shape, and the distortion of flats.  We show analytic validation of the diffusion physics, reproduce two kinds of edge distortion, and show qualitative validation of field-free regions, lithography errors, and fringing.  We also demonstrate that there are two related effects of doping variation having different observable consequences.  We show that field distortions from accumulated electrons lead to intensity-dependent point-spread-functions and the sub-linear variance in flats.  The method is implemented in the Photon Simulator (PhoSim) and the code is publically available.
\end{abstract}

\keywords{telescopes-- instrumentation: detectors--}

\section{Introduction}

Simulations have become increasingly more important in interpreting X-ray (\citealt{peterson2004}; \citealt{peterson2007}; \citealt{andersson2007}; \citealt{davis2012}) and optical astronomical observations (\citealt{ackermann2012}; \citealt{lane}, \citealt{ellerbroek}; \citealt{lelouarn}; \citealt{britton}; \citealt{jolissaint}; \citealt{bertin}; \citealt{dobke}; \citealt{peterson2015}, \citealt{rowe2015}).  The considerable complexity of the atmosphere, telescope, and camera imprint many systematics on the measurements of astronomical objects.  Some recent simulators use parameterized models of  point-spread-functions to generate synthetic images (\citealt{bertin}; \citealt{dobke}; \citealt{rowe2015}).    In our previous work (\citealt{peterson2015}, \citealt{peterson2019}, \citealt{burke2019}), however, we developed a ab initio physics simulator that tracks photons and the subsequent converted electrons through the atmosphere, telescope, and camera called the Photon Simulator (PhoSim).  PhoSim is publically available at:  https://bitbucket.org/phosim/phosim\_release/wiki/Home.
It is important to represent both the photon and electron interactions as well as the interacting components (atmosphere, telescope, and camera) using appropriate physics.  In \cite{peterson2019}, we implemented the complete physics describing the deformation of the optics to expand that work.  Another important area of simulation is the individual sensors in a camera.
In particular, in \citealt{peterson2015}, we noted that distortion of electric field lines from their ideal shape were important but did not describe how to represent them.  These distortions of electric field and other complications and how they are represented using formal Photon Monte Carlo methodology are described in this work.

Non-ideal sensor effects form some of the most difficult effects to properly calibrate and mitigate (\citealt{downing2006}; \citealt{stubbs2014};\citealt{antilogus2014}; \citealt{plazas2014}; \citealt{gruen2015}; \citealt{tyson2015}, \citealt{astier2015}, \citealt{rasmussen2015}, \citealt{magnier2018}); \citealt{astier2019}).  The point spread function (PSF) size and shape, the differential astrometric pattern, and the differential photometric pattern are all affected by the physics of non-ideal sensor features.  These systematic errors tend to be hard to remove, because since the sensor is, by definition, in the image plane.  Because of that, the effects can affect every source or every pixel or even part of a pixel differently.  The non-flatness of the sensor, the distortion of parallel field lines, contamination, and other imperfections all result in complex systematics in the PSF, astrometry, and photometry calibrations.  Most of these effects depend on the photon wavelength as well as the source location, since the conversion depth is a strong function of wavelength.  Below, we demonstrate that a proper way to understand these effects and eventually calibrate them is to follow a Photon Monte Carlo approach with the appropriate approximations of electrostatic physics describing the sensor.  We are also particularly interested in coupling the sensor distortions in an efficient numerical approach, so we can simulate large surveys.

The paper is organized as follows.  In \S2, we describe the implementation of non-ideal sensor physics in the PhoSim code using Photon Monte Carlo techniques.  In \S3, we then describe the observational effects of the sensor physics.

\section{Sensor Distortions}

In this work, we are concerned with the implementation of non-ideal aspects of sensors that ultimately impact astronomical measurements.  For simplicity, we assume it is a modern charge-coupled device (CCD) constructed out of a rectangular slab of Silicon (see \citealt{janesick1992}, \citealt{holland2003}).  Some of this work can be generalized to different materials and readout schemes as described in \cite{burke2019}.  We focus on non-ideal details which appear endemic to all modern sensors, and generally ignore the various ways that sensors could operate ineffectively but generally not common to all devices.  Some of these will be more important for certain applications.  In particular, some of these details are more relevant for thicker ($>50$ microns) devices and others for thinner ($<50$ microns) devices, so this parameter is studied in more detail in \S3.
Throughout this paper, we refer to the dimension parallel to the intended photon propagation as the $z$-axis and the other two dimensions as $x$ and $y$.  In the simulation, the position of the sensor can be offset in all three dimensions ($dx$, $dy$, and $dz$) and the sensor can be rotated in all three dimensions from the ideal orientation.  All of these perturbations simply affect the intercepts of the incoming photons.  In addition, the surfaces describing the rectangular slab can be distorted from their ideal flat shape.  We represent this distortion by a series of polynomials, either Zernike or Chebyshev, that locally describe the surface at a given location, $z(x, y)$.  The geometry and shape have an effect on the various physics that we describe below.

\subsection{Photon Propagation and Conversion}

Within the Silicon, we model the path of the photon as described in~\cite{peterson2015}.  This includes the wavelength-dependent refraction, which alters the ray trajectories.  The temperature-dependent mean free path is calculated and the photon is propagated along a path based on a random realization of the mean free path.  There are two complications to the photon conversion.  One is that interference (fringing) occurs and modifies the reflection probabilities from the surfaces.  The second is that the photon may convert in a region where the electron cannot reach the readout and will likely be reabsorbed.  Note that throughout this paper, for simplicity we assume that the photon converts into an electron rather than a hole, but there is no loss of generality and the expressions for electron mobility are replaced by hole mobility in hole-producing devices in PhoSim.

As described in \cite{peterson2015}, we implement fringing by performing a one-dimensional EM calculation if the photon reaches the back surface.  We expand that calculation to consider multiple relective layers of the sensor.  We include both bare Silicon in the slab as well as an oxide layer.  Then the reflectivities of the three boundaries are given by

    $$\rho_1 = \frac{\cos^p{\theta_v} - n_{Si} \cos^p{\theta_{Si}}}{\cos^p{\theta_v} + n_{Si} \cos^p{\theta_{Si}}}$$
    $$\rho_2 = \frac{n_{Si} \cos^p{\theta_{Si}} - n_{Ox} \cos^p{\theta_{Ox}}}{n_{Si} \cos^p{\theta_{Si}} + n_{Ox} \cos^p {\theta_{Ox}}}$$
    $$\rho_3 = \frac{n_{Ox} \cos^p{\theta_{Ox}} - \cos^p{\theta_v}}{n_{Ox} \cos^p{\theta_{Ox}} + \cos^p{\theta_v}}$$

\noindent
where $\theta_v$, $\theta_{Si}$, and $\theta_{Ox}$ are the incidence angles in the the vacuum, Silicon, and oxide layer, respectively, $p$ is the polarization (represented as either +1 or -1), and $n_{Si}$ and $n_{Ox}$ are the wavelength-dependent indices of refraction.  Then the reflection ratios, $\Gamma_1$ and $\Gamma_2$ are given by

    $$\Gamma_2 = \frac{\rho_2 + \rho_3 e^{-2 i \delta_2}}{1 + \rho_2 \rho_3 e^{-2 i \delta_2}}$$
    $$\Gamma_1 = \frac{\rho_1 + \gamma_2 e^{-2 i \delta_1}}{1 + \rho_1 \gamma_2 e^{-2 i \delta_1}}$$

\noindent
where $\delta_1$ and $\delta_2$ are given by

    $$\delta_1 = \frac{2 \pi}{\lambda} t_{Si} n_{Si} \sqrt{1.0 - \frac{\sin^2{\theta_v}}{n_{Si}^2}}$$
    $$\delta_2 = \frac{2 \pi}{\lambda} t_{Ox} n_{Ox} \sqrt{1.0 - \frac{\sin^2{\theta_v}}{n_{Ox}^2}}$$

\noindent
where $t_{Si}$ and $t_{Ox}$ are the local layer thicknesses.   Then the reflection probability in one segment is given by $\Gamma_1^2 + 1 - \rho_3^2$.  If the photon is not converted in the distance it is predicted to travel via the photon conversion calculation, then we iteratively repeat the calculation to simulate multiple reflections.

In many devices, particularly back-illuminated ones, there may be a field-free region above the depleted volume.  Thus, photons that convert in the field-free region do not ever result in a photoelectron reaching the readout.  The field-free region can have a complicated spatial structure depending on how the Silicon is thinned for back illuminated devices.  Laser annealing techniques in particular result in a structured non-uniform pattern that follows the cadence of the rectangular laser annealing beam's footprint.  It is straight-forward to represent this in a Monte Carlo method.  We simply define a volume in three-dimensions according to a likely cadence of the beam's footprint where there is no field.  Then if the photon converts in this volume, we remove the electron from the simulation.  We therefore assume that the electron does not reach the readout in order to reproduce the spatial variation, however, in real devices it is possible that the electron could reach the readout in other pixels depending on the exact field structure.  This naturally results in the complex wavelength-dependence, since shorter wavelength photons are much more likely to convert in this relatively small volume.

To model this, we consider the path of a laser annealing footprint tilted at an angle, $\theta$, with respect to the pixel grid, so that pixel positions are multiplied by a 2x2 rotation matrix.  We then take the modulus of the rotated pixel positions with the footprint lengths ($x_L$, $y_L$) with a subtracted overlap ($x_o$, $y_o$).   Then if the pixel is within the footprint length, we define the dead layer depth as $d$.  If the pixel is within one overlap region, we double the depth and if it is within two of the overlap region we triple the depths.  We repeat the whole process for multiple passes of the annealing process.  The result of this, is that there is a small field-free region at the top of the sensor with a complex ``brick wall'' pattern as we show in \S3.  We then destroy electrons that are generated in the field free region.

\subsection{Electrostatic Simulations and Electron Diffusion}

In \cite{peterson2015}, the vertical component of the electric field was calculated as

$$ E_z (z) = \frac{V}{t_{Si}} + \frac{e}{\epsilon_{Si}} \int_{z_{coll}}^z dz n_d(z)$$

\noindent
where $V$ is applied potential, $t_{Si}$ is the thickness, $\epsilon_{Si}$ is the permittivity in Silicon,  $e$ is the electron charge, $n_d$ is the doping density.  We pre-calculate this at various numerical depths in the sensor slab.

The diffusion is then given by the $\sqrt{ 2 \mu kT / e t_{coll}}$ where $\mu$ is the electron mobility which is a function of temperature and electric field, $k$ is Boltzmann's constant,  $T$ is the temperature, and $e$ is the electron charge.  The collection time, $t_{coll}$, is given by another piecewise integral,

$$ \int_{z_{coll}}^{z} \frac{dz}{\mu E_z(z)} $$

To represent some non-ideal sensor properties, the one dimensional solution above is insufficient.  Instead, we have to determine the full three-dimensional electrostatic solution.  To implement this numerically, we consider the electrostatic potential in a three-dimensional cartesian coordinate system, $\phi(x,y,z)$.  We then solve Poisson's electrostatic equation,

$$ \nabla^2 \phi = -\frac{\rho}{\epsilon_{Si}}$$

\noindent
where $\rho$ is the charge density, and $\epsilon_{Si}$ is the permittivity.  We construct the slab of Silicon consisting of a series of rectangular volume elements and allow for a different width in $dx$ and $dy$ than the vertical element in $dz$.  We then make the total number of elements equal to $N$ in each dimension and $N^3$ in the overall volume.

To solve the equation numerically, we randomly choose a point in the three dimensional volume.  We then evaluate the accuracy of the Poisson equation by first calculating the numerical derivative in each of the three dimensions and calculate the differential between that and the charge density divided by the permittivity as

$$ e_0 = \vline \frac{\phi(x+dx,y,z)+\phi(x-dx,y,z)-2 \phi(x,y,z)}{dx^2} + $$
$$     \frac{\phi(x,y+dy,z)+\phi(x,y-dy,z)-2 \phi(x,y,z)}{dy^2} +$$
$$    \frac{\phi(x,y,z+dz)+\phi(x,y,z-dz)-2 \phi(x,y,z)}{dz^2} + \frac{\rho(x,y,z)}{\epsilon} \vline $$

\noindent
We then evaluate the error, $e_+$, when a small amount, $\delta$ is added to $\phi(x,y,z)$.  We also evaluate the error, $e_-$, when a small amount is subtracted from $\phi(x,y,z)$.  We then modify $\phi$ according to a double Metropolis-Hastings (\citealt{metropolis1953}, \citealt{hastings1970}) criterion.  If $e_+$ is less than $e_-$ then we accept $e_+$ if it is less than $e_0$ or if a uniform random deviate is less than $e^{-\frac{e_+-e_0}{\alpha}}$.  Similarly, if $e_-$ is less than $e_+$ then we accept $e_-$ if it is less than $e_0$ or if a uniform random deviate is less than $e^{-\frac{e_--e_0}{\alpha}}$.  After choosing all of the points in the lattice, we set the value of $\delta$ to be equal to $10^{-6 \frac{i}{m_i}}$ where $i$ is the iteration, and $m_i$ is the maximum iteration.  We set the value of $\alpha$ to be equal to a fraction $f$ of the average current error.  We achieved robust results with $m_i$ of 2000, and $f$ set to 0.1.

For the boundaries, we consider both Dirichlet and Neumann conditions for the six surfaces.  The Dirichlet conditions are accomplished by simply setting it to a given constant potential on the top and bottom surfaces.  For the Neumann boundary conditions, if the surface is in the $y$-$z$ plane, then a point on the surface is given by

$$ \phi(x,y,z) = \phi(x+dx,y,z) - \frac{\sigma}{\epsilon} dx$$

\noindent
where $\sigma$ is the surface charge, and for the other surfaces the coordinates are exchanged.  We then iterate the value of the potential on all points where the boundary conditions do not apply.

To compute the electric field, we evaluate the numerical derivative of the electric potential as

$$ E_x  = \frac{\phi(x+dx,y,z) - \phi(x-dx,y,z)}{2 dx} $$
$$ E_y  = \frac{\phi(x,y+dy,z) - \phi(x,y-dy,z)}{2 dy} $$
$$ E_z  = \frac{\phi(x,y,z+dz) - \phi(x,y,z-dz)}{2 dz} $$

This electric field can then be used in the calculation of the electron path.

As discussed in \citealt{peterson2015}, the lateral shift of the electron due to a transverse field is given by

$$\Delta x=\int dz \frac{E_x(x,y,z)}{E_z(x,y,z)} $$
$$\Delta y=\int dz \frac{E_y(x,y,z)}{E_z(x,y,z)} $$

\noindent
where the integral should be evaluated from the conversion point to the collection surface.  Similarly, the diffusion discussed above can be calculated based on an integral at the photo-electron conversion point.  However, if the electron drifts laterally by a significant amount (a pixel) either from diffusion or lateral fields, then it makes it difficult to evaluate this integral for all cases.  This only occurs for significant changes to the field profile that occurs within a pixel (e.g. see the discussion on accumulated charges below), but is not significant for many details.  In order to evaluate this accurately for all cases, it is straightforward to break up both the diffusion integral as well as the lateral shift integrals above in a piece-wise manner.  Then, the path of the electron is broken into a series of vertical segments where both the lateral shift for that segment as well as the overall diffusion for that segment is used to predict a new $(x,y)$ position.  The segmentation of the calculation, however, does not have to represent the real random walk of the electron.  In practice, splitting the vertical slices into 4 segments achieves essentially the same numerical accuracy as an arbitrarily large number.  Ultimately, the accuracy depends on the inhomogeneity of the electric field which we discuss in the following sections.

\subsection{Edge Distortion}

The simplified 1-D calculation of the electric field assumes that we have an infinite slab of Silicon and therefore no distortion near the edge of the Silicon.  However, in various device designs is likely that the electric field is distorted near the edge where it will be less constrained by the conducting surfaces.  This can be simulated as a surface charge on the edge surfaces that are not held at constant potential.  If we add a surface charge, $\sigma$, on the edge surfaces using the Neumann condition, we find the electric field can be parameterized by

$$ \vline \vec{E} \vline = e^{-\frac{x n_d}{\sigma}}$$

\noindent
where $n_d$ is the average bulk doping density and $x$ is the distance from the edge.  Thus, the lateral field decays from the edge.  The direction of the field is perpendicular to the edge and we use two components in the corners of the device.

\subsection{Doping Variation}

In CCD manufacturing when the Silicon boule is formed that is eventually used to make the individual sensor wafers, the molten Silicon is mixed with a Boron or Phosphorus dopant and then it crystallizes as it is cooled while rotating.  The dopants, however, have a different segregation coefficient.   If the thermal axis (axis of thermal symmetry) is not perfectly aligned with the rotational axis then the dopant while not be distributed uniformly.  Since the Silicon boule is continually rotated, the non-uniformity will tend to have a ring-like structure as it cools radially.  This leads to a ``tree-ring'' like variation of the dopant.

The rings structure is complex since it depends on interplay between the crystallization, rotation, and the different materials.  Therefore the pattern can only be inferred indirectly (as discussed in \S3).  However, given a parametrized doping pattern we can correctly predict the response to the electric field and then follow the response to the electrons.  We found that empirically we can describe a range of doping variations by

$$ n(r) = n_0 \left( 1 +  \sum_{i=1}^N \frac{\alpha}{\sqrt{N}} \sin{ \left( \frac{2 \pi r}{P_i e^{-r/r_0}} + \phi_i \right) }  \right) $$

\noindent
where $P_i$ is the period, $\phi_i$ is the phase, $r_0$ attenuates the period, $n_0$ is the nominal doping density, and $\alpha$ is the overall amplitude.  The multiple components are used to empirically match realistic doping variations.  The individual period is chosen from a uniform distribution between a lower period and a higher period.  The amplitude is approximately constant, but can also be further described by a function of form

$$ \alpha = t_1 + \frac{1-t_1}{1.0 + e^{-\frac{r - r_r}{t_r}}}$$

\noindent
where $t_1$, $t_r$, and $r_r$ are empirical parameters.

The dopant variation has two important consequences.  One is that it creates a lateral field having a sinusoidal variation between the two surfaces.  A second consequence is that the overall level of charge diffusion will also vary because the strength of the vertical field is affected by the level of doping as described in \S2.2.  For the first effect, we found an electric solution from the three dimensional Poisson simulation above by including a sinusoidal variation in dopant density (uniform in the z-direction).  The amplitude of the lateral field is given by

$$ A = \frac{dn}{dx} \frac{n}{<n>} t_{Si} <P_i> \frac{e}{(2 \pi)^2 \epsilon_{Si}}$$

\noindent
where $\frac{dn}{dx}$ is the numerical derivative of the doping map, $\frac{n}{<n>}$ is the relative amplitude of the doping map, $P_i$ is the average local period, $t_{Si}$ is the device thickness, $e$ is the electric charge, and $\epsilon_{Si}$ is the permittivity.

For the second effect, we simply store the dopant density at all $(x,y)$ and then use a locally different diffusion estimate.  The combined result of both effects is then complex, because the diffusion distortion and lateral shifts are occuring simultaneously during the simulation.  We can separate the observable consequences of each effect, since we can turn each on and off as described in \S3.

\subsection{Accumulated Charges}

It has been noted that the electrons collected in pixel can significantly alter the local electric field (\citealt{downing2006}, \citealt{rasmussen2015}, \citealt{astier2019}).  This then affects the final location of subsequent electrons.  This has a number of observable consequences that we explore in \S3.  To simulate this, we performed various electrostatic simulations with the three dimensional method described above.  We performed a small-scale simulation where a charge was placed at the collection surface.  We then solved the electrostatic equation and studied the resulting field.  We found that the magnitude of the perturbation of the electric field due to the accumulated charge can be described by a diluted Coulomb like solution as

$$ | \vec{E} | = \frac{f e}{4 \pi \epsilon_{Si} ( z^2 + r^2 )} $$

\noindent
where $f$ is the dilution factor that is determined numerically.  Note that this superposition solution is not perfect (i.e. $f \neq 1$) because the field still has to satisify the boundary conditions.  However, we are mainly interested in having an approximate solution, because the goal is simply to predict which of the subsequent electrons will jump to a different pixel than they would have without the accumulated charge existing.  For large values of $r$, this solution is not appropriate and the field will exponentially decay (\citealt{pumplin1969}), but for the adjacent pixels where $r$ is a small fraction of the sensor thickness this approximation is valid.

An additional detail that is important to include is that the accumulated charge does not occupy a single point at the center of the pixel.  To implement this, we include two parameters that describe the standard deviation of the charge in the potential well, $\sigma_x$ and $\sigma_y$.  Then the pixel shift due to an accumulated charge is given by first computing the function as in \cite{peterson2015} as

$$ f_i(r, z) = \int_{z-dz}^{z+dz} \frac{E_i(r,z)}{E_z(r,z)} dz $$

\noindent
where $i$ is either the $x$ or $y$ component, $dz$ is a numerical step that computes the local kick at each segment of the sensor slab, and $r$ is the radial distance.  Then the lateral kick is given by evaluating this function

$$ \delta X_i = Z f_i(\sqrt{r^2 + \frac{x^2}{r^2} \sigma_x^2 + \frac{y^2}{r^2} \sigma_y^2}, z)$$

\noindent
where $Z$ is the current accumulated charge.  Then, we evaluate this kick for not just the accumulated charge in the pixel where the electron is currently, but we also perform the same calculation at all the neighboring pixels.  In practice, most observable consequences are captured by evaluating the neighbors in a 3x3 grid.  However, some correlations in bright flats would require a larger grid of pixels.  The advantage of this calculation is that it can be done efficiently on a photon by photon basis.  The field will grow as the charges start accumulating in the simulation, but the calculation of $f$ can be done on a dense numerical grid once in the entire simulation.  This makes the Monte Carlo of this complex detail both efficient and accurate.  Charge stops that prevent the electrons from bleeding into neighboring rows are implemented as a line charge with the same formalism, but do not have a large effect on final electron locations.

\subsection{Lithography Errors}

The pixels in a sensor may not be perfectly placed in a geometric grid.  Most likely, this can be represented in errors in both column locations and row locations.  To implement this, we place a random error in the physical size of the final pixels in both the row and the columns.  Then when the electron reaches the collection plane, its final pixel is determined by the perturbed pixel grid that is modified by the random lithography errors.

\section{Sensor Observational Effects}

There are a number of observational consequences to the implemented physics described in \S2.  In many cases, there are actually multiple observational consequences to a single physical effect.  Following, we demonstrate the observational effects and describe various validation calculations.

\subsection{Photon and Electron Interactions}

\begin{figure*}[htb!]
\begin{center}
\includegraphics[width=1.57\columnwidth]{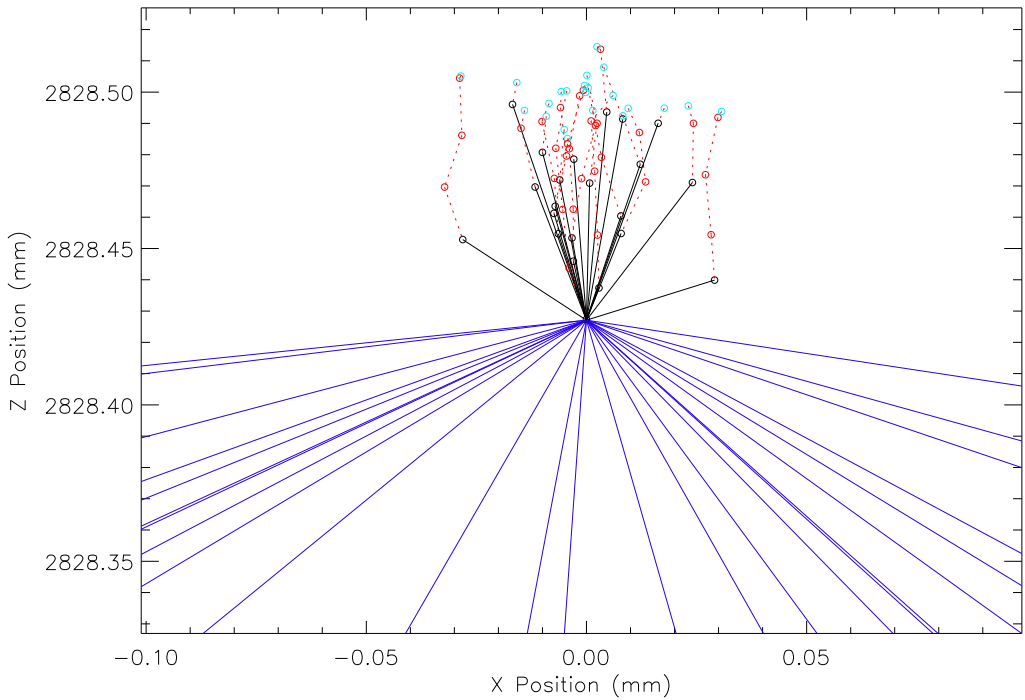}
\end{center}
\caption{\label{fig:label1} Basic photon and electron interactions in the Simulation.  The photons are propagated from the bottom of the figure in a cone following the f/\# of our generic telescope (purple).  The interaction with the silicon occurs at the convergence point, and a narrower cone due to the index of refraction of the Silicon is shown in black rays.  Photon conversions into electrons are shown by the red dashed lines.  The electrons then diffuse in numerical segments until they reach the readout plane (light blue).  Note this plane is tilted from our viewpoint.}
\end{figure*}

In Figure~\ref{fig:label1}, we show the overall basic interactions in the simulation by simulating a dozen of photons.  The path of the photons before reaching the silicon and then the much narrower cone due to refraction of the photons while propagating in the Silicon is shown.  Then the photons propagate for a random amount according to the photon conversion physics.  After interaction the electrons are propagated in segments to represent the variable diffusion at each height in the Silicon.  Finally, the electrons reach the readout plane at the top of Figure~\ref{fig:label1}.

\subsection{Diffusion Validation}

\begin{figure*}[htb]
\begin{center}
\includegraphics[width=1.57\columnwidth]{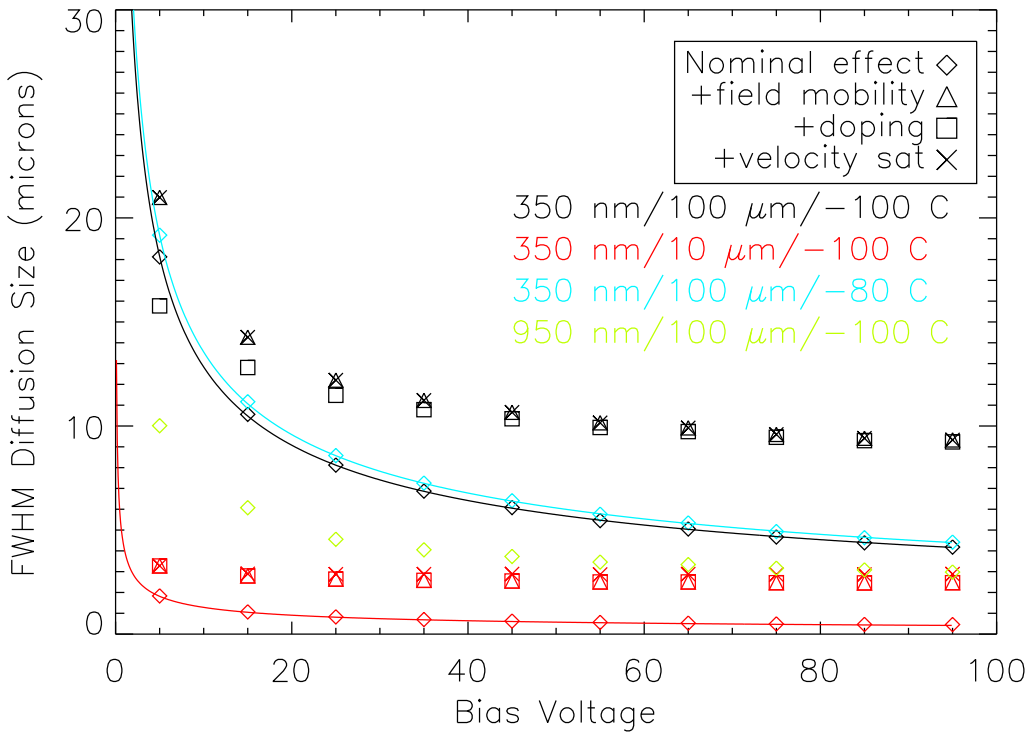}
\end{center}
\caption{\label{fig:label2} The full width at half maximum of the electric charge diffusion as the electrons reach the readout in various configurations.  The diffusion size vs. bias voltage is shown for:  a simulation of 350 nm light through a 100 $\mu$m thick device at -100 Celsius (black), 350 nm light through at 10 $\mu$m thick device at -100 Celsius (red), 350 nm light through a 100 $\mu$m thick device at -80 Celsius (blue), and 950 nm light through a 100 $\mu$m thick device at -100 Celsius (green).  The different symbols correspond with levels of physics fidelity.  The diamonds ignore the effects of:  1) the doping, 2) the velocity saturation, 3) and the decrease in mobility due to the vertical electric field strength.  The triangles add the decrease in mobility due to the field strength, the squares add the effect of the doping, and the cross adds the effect of the velocity saturation.  The lines show the analytic prediction in the limit of short wavelength light.}
\end{figure*}

We evaluate the accuracy of the diffusion calculation in Figure~\ref{fig:label2}.  We first simulate a set of photons as in Figure 1.  We set the index of refraction in the Silicon effectively to infinity, so the photons are incident normal to the Silicon.  This enables us to isolate the effect of diffusion.  We perform four sets of simulations:  1) photons with wavelength of 350 nm and a 100 micron detector at -100 Celcius, 2) photons with wavelengths of 350 nm and a 10 micron detector at -100 Celcius, 3) photons with wavelength of 950 nm and a 100 micron detector at -100 Celcius, and 4) photons with wavelengths of 350 nm and a 100 micron detector at -80 Celcius.  With these simulations, we also turned off some of the physics to validate the individual parts.  To compare the diffusion calculation with an analytic prediction, we ignored 1) the effect of the doping on the electric field profile, 2) the electric field dependence of the mobility by setting it to the field-free value, and 3) the effect of velocity saturation on the electron mobility.  Then the standard deviation of the lateral diffusion is given by

$$ \sigma = \sqrt{\frac{2 k T}{e V}} t $$

\noindent
where $V$ is the bias voltage, $T$ is the temperature, $k$ is Boltzmann's constant, $e$ is the electron charge, and $t$ is the sensor thickness.  The four simulations show perfect agreement with the analytic prediction at all bias voltages.  By using the electric field dependence of the mobility the overall diffusion is increased significantly as expected since the mobility vertically decreases which increase the collection time.  The effect of doping is visible at low bias voltages for the 100 micron simulations, since the overall electric field profile will be most influenced when $\frac{V}{d}$ is small.   Conversely, the effect of the velocity saturation is most visible at high bias voltage for the 10 $\mu m$ simulation, since the velocity of the electron will only approach the saturation velocity when $\frac{V}{d}$ is large.

\subsection{Edge Distortion Effects}

\begin{figure*}[htb]
\begin{center}
\includegraphics[width=1.57\columnwidth]{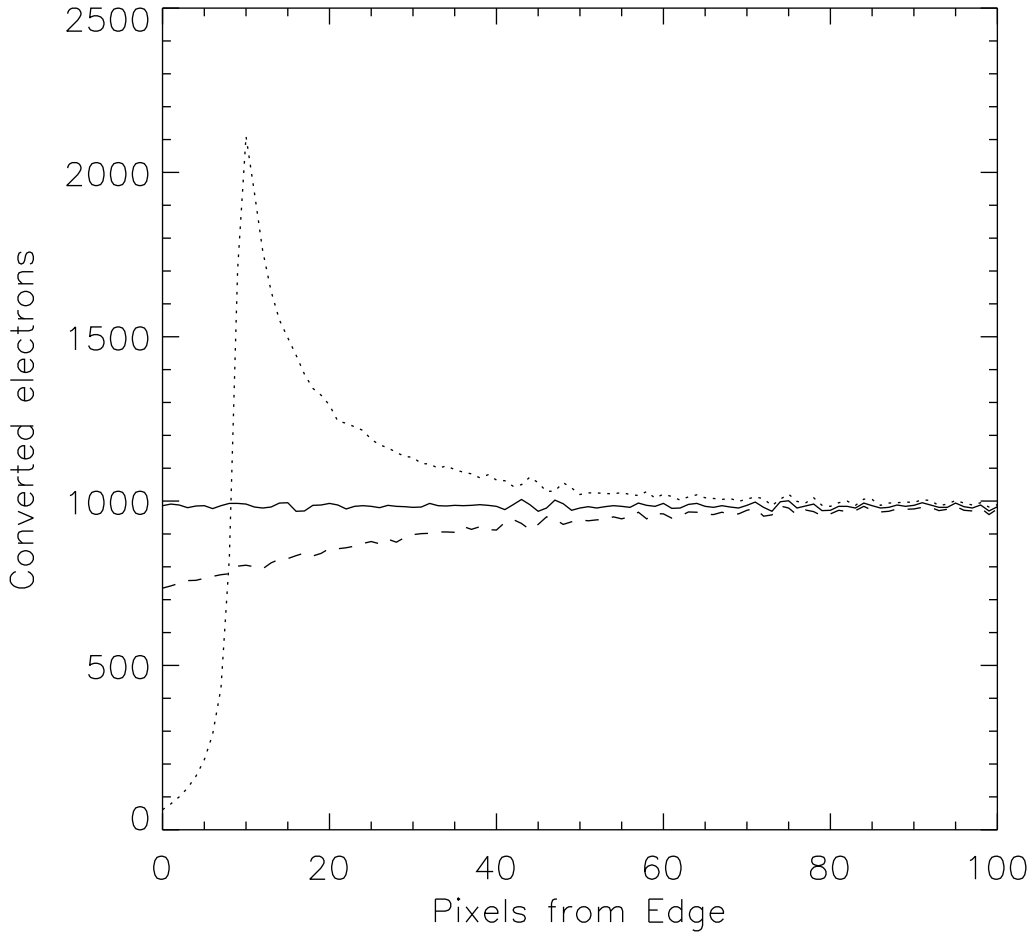}
\end{center}
\caption{\label{fig:label3} The edge roll off of a simulated flat with 3 different values of the surface charge.  The  solid, dotted, and dashed lines correspond with surface charge of 0, $-2 \times 10^{10}$, and $2 \times 10^{10}$.}
\end{figure*}

The effect of the edge distortion can be seen by simulating the effect of different effective surface charge as seen by two sets of simulations.  We set the surface charge of $2 \times 10^{10}$ and $-2 \times 10^{10}$ electrons $\mbox{cm}^{-2}$ and generate a flat where photons are produced at a range of wavelengths and at all incident angles.  The negative surface charge implies there is repulsive force near the boundary whereas a positive surface charge is an attractive force due to the net electric field outside the volume.  The decrease in flux near the edge is then seen in Figure~\ref{fig:label3}.  The effect of positive surface charge leads to a gradual roll off near the edge, whereas the effect of negative surface leads to an excess of flux several pixels near the edge and an asymmetric fall off on either side.  The simplicity of the positive charge fall off is simply due to the electrons being pulled away from the imaging Silicon, whereas the asymmetric pattern of the negative surface charge is due to how many electrons can be pushed from the non-imaging area and differing amounts.  The latter effect was qualitatively observed in \cite{estrada2010}.

\begin{figure*}[htb]
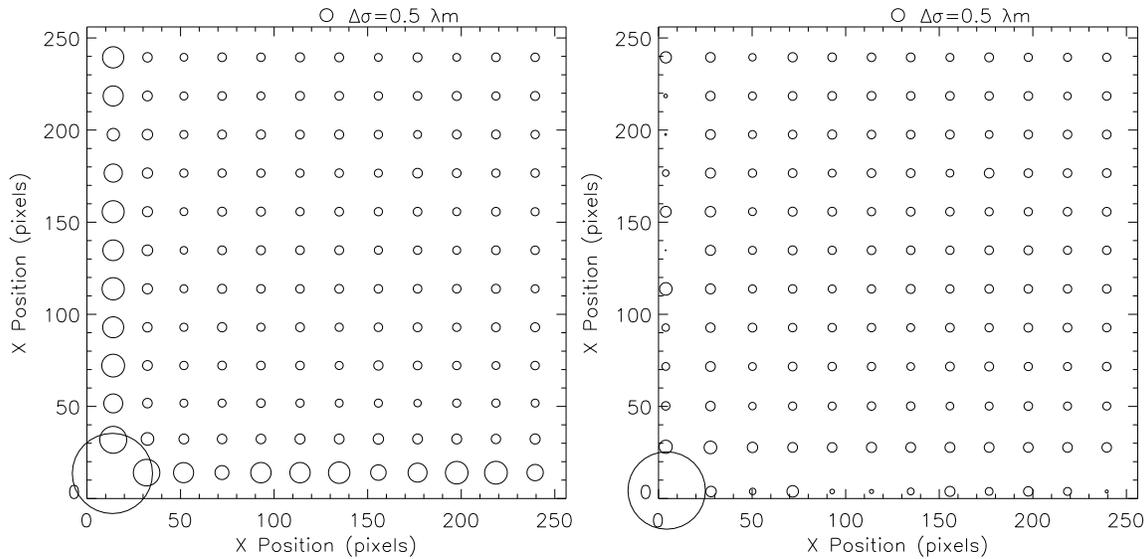

\begin{center}
\includegraphics[width=0.87\columnwidth]{f06.eps}
\includegraphics[width=0.87\columnwidth]{f09.eps}
\end{center}
\caption{\label{fig:label4} The PSF size of a grid of stars near the corner of the sensor.  The left plot shows a surface charge of $2 \times 10^{10}$ and the right plot shows a surface charge of $-2 \times 10^{10}$.  In both cases, the diffusion pattern becomes modified towards the edge.}
\end{figure*}

\begin{figure*}[htb]
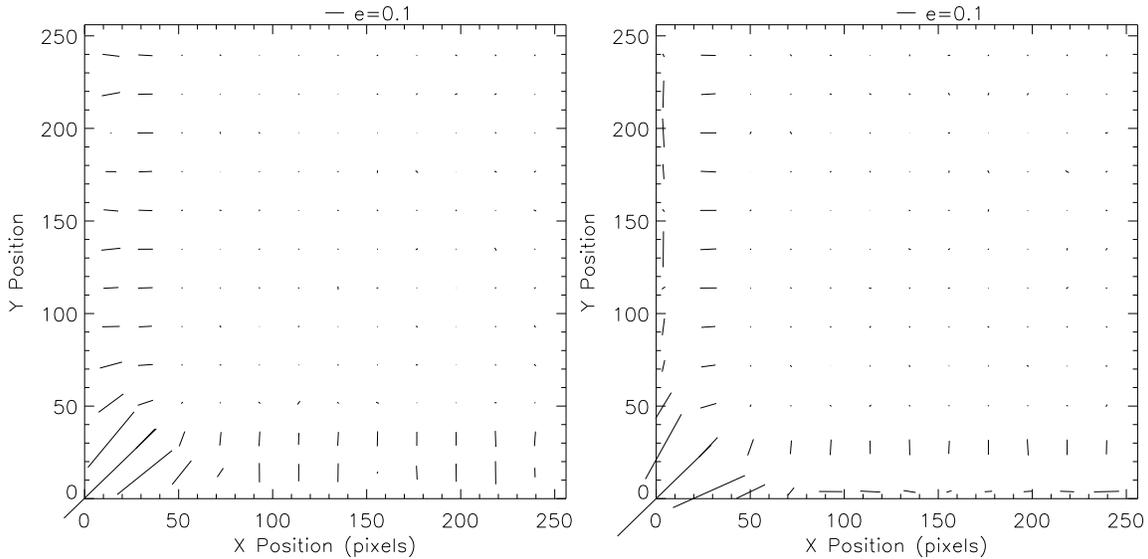

\begin{center}
\includegraphics[width=0.87\columnwidth]{f04.eps}
\includegraphics[width=0.87\columnwidth]{f07.eps}
\end{center}
\caption{\label{fig:label5} The PSF ellipticity of a grid of stars near the corner of the sensor.  The left plot shows a surface charge of $2 \times 10^{10}$ and the right plot shows a surface charge of $-2 \times 10^{10}$.  In both case, the ellipticity is tangent to the edge due to the charge being pushed or pulled from the edge.}
\end{figure*}

\begin{figure*}[htb]
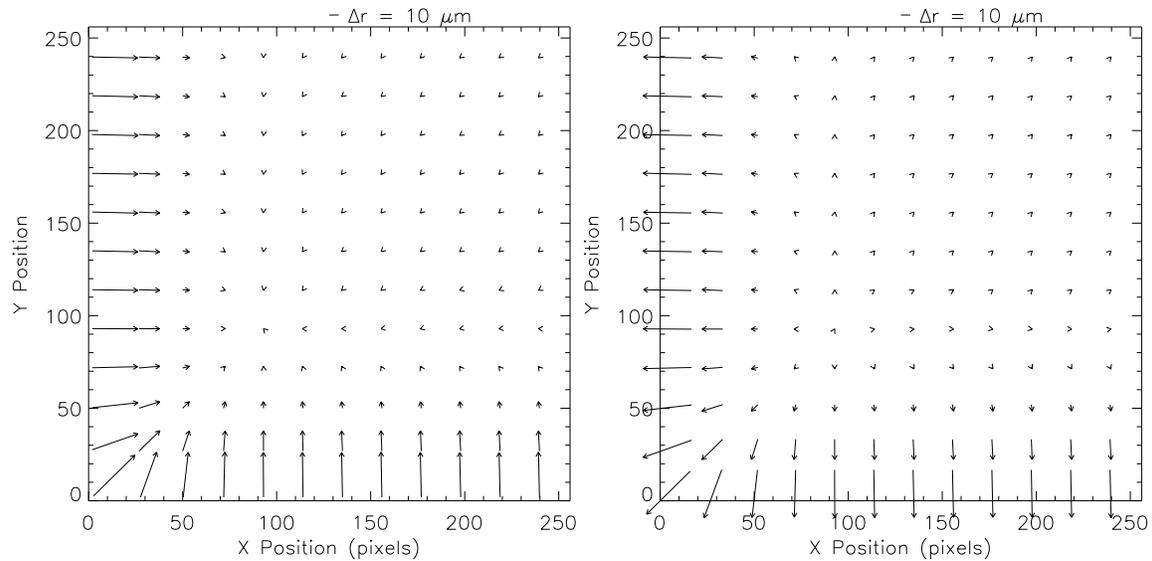

\begin{center}
\includegraphics[width=0.87\columnwidth]{f05.eps}
\includegraphics[width=0.87\columnwidth]{f08.eps}
\end{center}
\caption{\label{fig:label6} The differential astronometry pattern of a grid of stars near the corner of the sensor.  The left plot shows a surface charge of $2 \times 10^{10}$ and the right plot shows a surface charge of $-2 \times 10^{10}$.  The left plot shows the charge being pushed several pixels from the edge, whereas the right part shows the charge being pulled from the edge.}
\end{figure*}

The effect of the edge roll off also has a variety of consequences in PSF patterns near the edge.  We simulate a grid of point sources across the edge and then measure their PSF properties.  Figure~\ref{fig:label4} shows the increase in PSF size near the edge.  Similarly, the PSF acquires an elliptical shape perpendicular to the edge in response to the pulling or pushing of the charge as shown in Figure~\ref{fig:label5}.  Finally, the astrometric position of the PSFs becomes distorted from the original grid as shown in Figure~\ref{fig:label6}.  The negative surface charge compresses the astrometric pattern, whereas the positive surface charge stretches the grid.  These results are consistent with \cite{plazas2014} edge observations with DECam that observed a 10$\%$ change of flat intensity near the edge and a corresponding 1.5$\mu m$ astrometric shift.  We observe a similar ratio with a 100$\%$ flat intensity change and a 25 $\mu m$ astrometric shift, so the surface charge would be consistent with a factor of 10 to 15 smaller implying a effective surface charge of approximately $2 \times 10^{9}$ electrons $\mbox{cm}^{-2}$.

\subsection{Qualitative Flat Variation}

\begin{figure*}[htb]
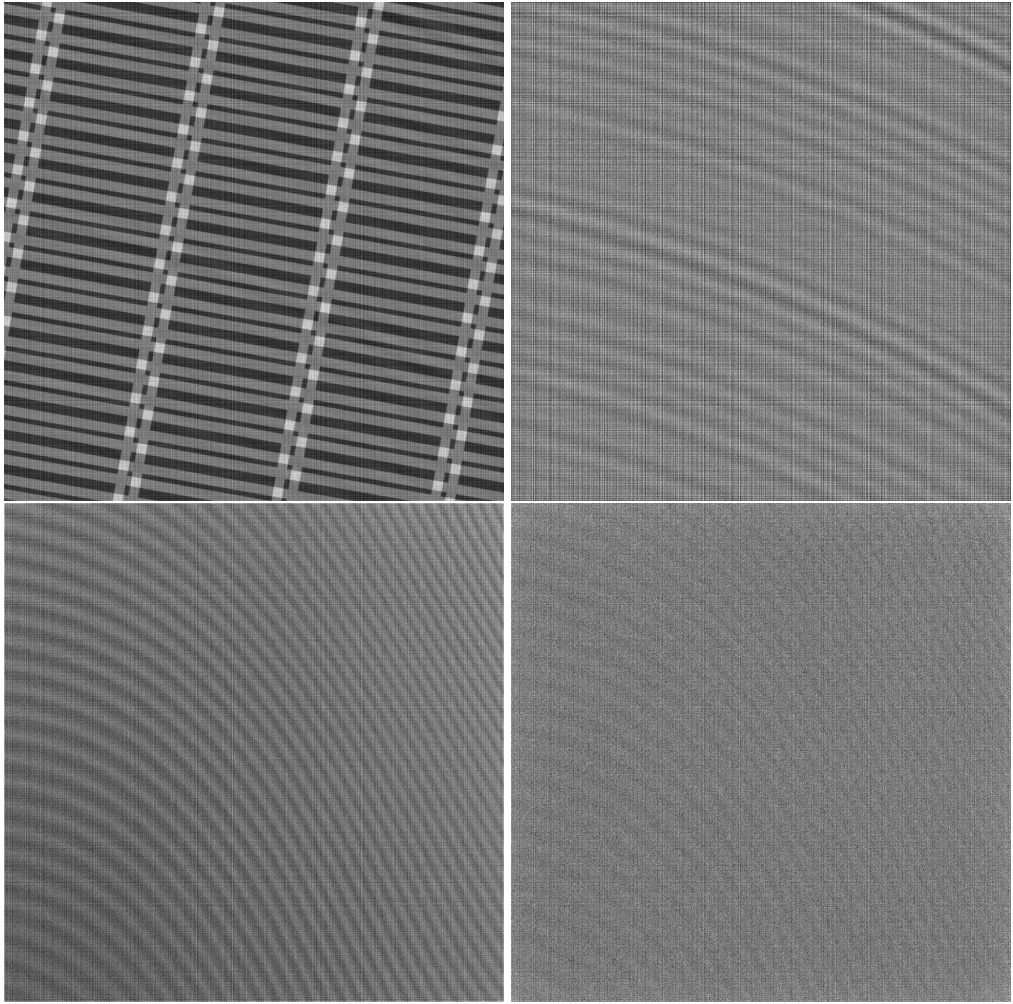

\begin{center}
\includegraphics[width=0.77\columnwidth]{f10.eps}
\includegraphics[width=0.77\columnwidth]{f11.eps}
\includegraphics[width=0.77\columnwidth]{f12.eps}
\includegraphics[width=0.77\columnwidth]{f23.eps}
\end{center}
\caption{\label{fig:label17} A series of monochromatic flats highlighting different aspects of the physics.  The flats are generated at 350, 500, 950, and 1000 (clockwise starting from top left).   The 350 nm flat shows the field free pattern.  The 500 nm flat shows the row/column lithography errors and the tree rings.  The 950 nm and 1000 nm show the fringing pattern.}
\end{figure*}

A number of effects are most visible by the modulation of a flat.  We perform a series simulation of large number of photons with a uniform set of angular incidence angles.  In order to evaluate the field free implementation described in \S2.4, we simulate a monochromatic flat at 350 nm.  At this wavelength, the mean free path of the photons in the Silicon is small and therefore the efficiency of the photon conversion will map the field free pattern.  In Figure~\ref{fig:label17}, we show the flat and the two layer overlapping brick wall pattern.  This is qualitatively similar to what is found in \cite{wei1998}.

The effect of row and column lithography errors can be see by simulating a flat at a mid-range wavelength.  In Figure~\ref{fig:label17}, we simulate a 500 nm monochromatic flat.  The column and row striping is visible with a 0.3\% variation in the x-direction and 0.2\% variation in the y-direction.  This is enough to make some electrons near the border shift to a neighboring pixel that they would not have otherwise.  Also visible in this image is the tree ring pattern that we discuss in the following section.

Finally, the effect of fringing can be visualized by simulating a long wavelength monochromatic flat.  In Figure~\ref{fig:label17}, we simulate a 950 nm monochromatic flat.  The interference pattern follows the variation in surface height.

\subsection{Two Tree Ring Effects}

\begin{figure*}[htb]
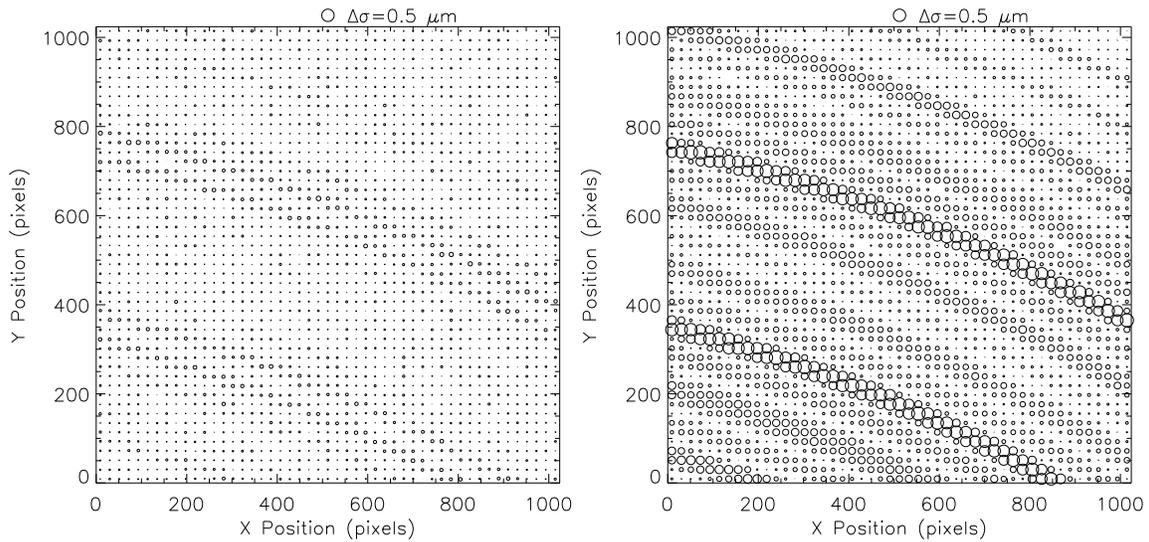

\begin{center}
\includegraphics[width=0.87\columnwidth]{f15.eps}
\includegraphics[width=0.87\columnwidth]{f18.eps}
\end{center}
\caption{\label{fig:label10} The PSF size of a grid of stars on a sensor with a tree ring doping pattern.  The left plot is the lateral field effect and the right plot is the vertical field effect.  The modification in PSF size is dominated by the vertical field effect.}
\end{figure*}

\begin{figure*}[htb]
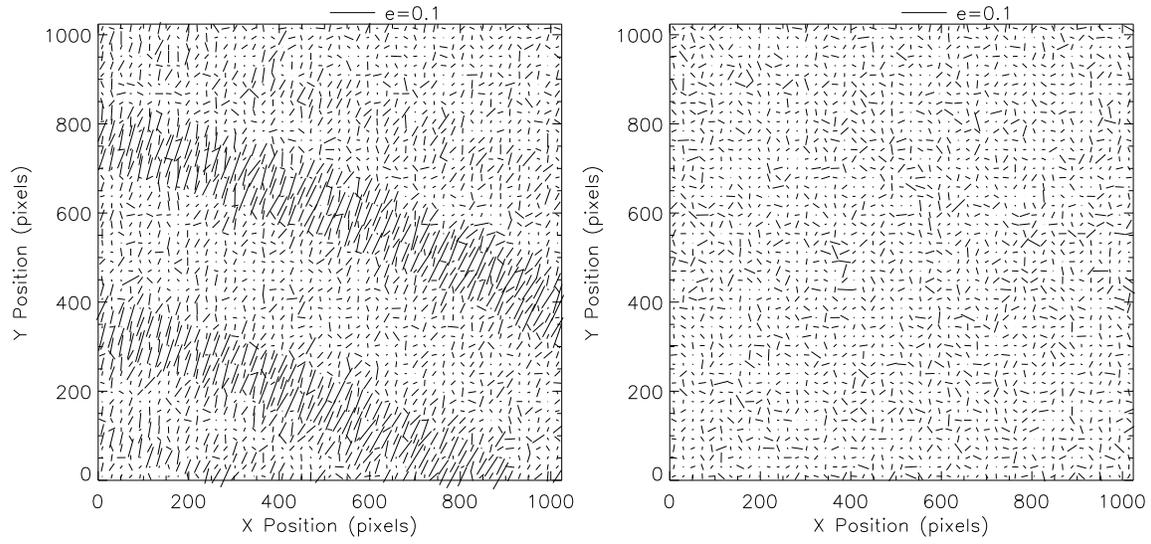

\begin{center}
\includegraphics[width=0.87\columnwidth]{f13.eps}
\includegraphics[width=0.87\columnwidth]{f16.eps}
\end{center}
\caption{\label{fig:label12} The PSF ellipticity of a grid of stars on a sensor with a tree ring doping pattern.  The left plot is the lateral field effect and the right plot is the vertical field effect.  The PSF ellipticity pattern is dominated by the lateral field effect.}
\end{figure*}

\begin{figure*}[htb]
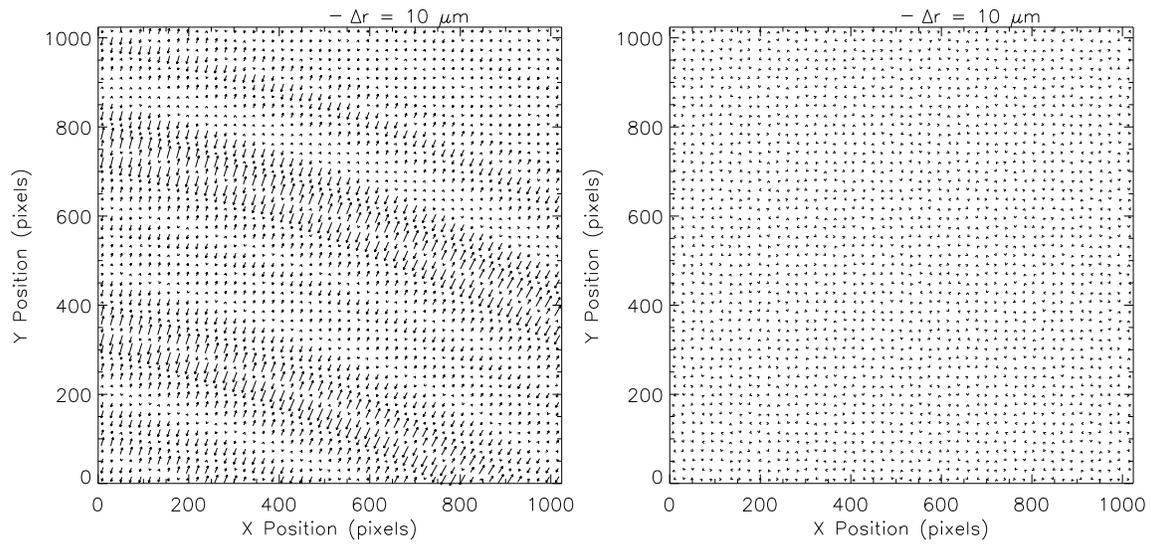

\begin{center}
\includegraphics[width=0.87\columnwidth]{f14.eps}
\includegraphics[width=0.87\columnwidth]{f17.eps}
\end{center}
\caption{\label{fig:label11} The differential astrometry pattern of a grid of stars on a sensor with a tree ring doping pattern.  The left plot is the lateral field effect and the right plot is the vertical field effect.  The astrometry pattern is dominated by the lateral field effect.}
\end{figure*}

\begin{figure*}[htb]
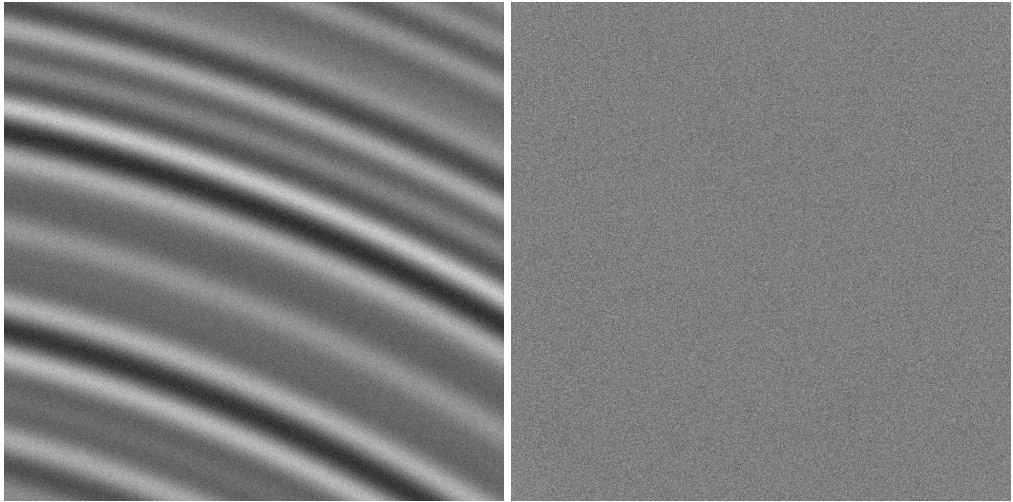

\begin{center}
\includegraphics[width=0.77\columnwidth]{f19.eps}
\includegraphics[width=0.77\columnwidth]{f20.eps}
\end{center}
\caption{\label{fig:label13} Simulation of a polychromatic flat of a sensor with a tree ring doping pattern.  The left plot is the lateral field effect and the right plot is the vertical field effect.  The flat variation is due to the lateral field effect.}
\end{figure*}

There are two physical effects that result from the doping variation present in devices:  1) the perturbed lateral electric field and 2) the modification of the diffusion by changing the vertical electric field profile.  \cite{magnier2018} hypothesized that the diffusion variation by the modification of the electric field could be an important contributor to PSF size variation.  We can determine the observable consequences by isolating each effect.  We exaggerate the effect for illustrative purposes by using a 50\% doping variation amplitude.  The average tree ring period is set to 2 mm with a variation of 0.4 mm.  In Figure~\ref{fig:label10}, we measure the effective size of the diffusion by simulating a grid of point sources.  The lateral effect does not show evidence of the tree ring pattern, whereas the vertical effect clearly shows the tree ring pattern.  In Figure~\ref{fig:label11}, we measure the astrometric displacement from the nominal grid.  The lateral effect shows the tree ring pattern, whereas the vertical effect does not.  Similarly, in Figure~\ref{fig:label12} we measure the ellipticity and show that the tree ring pattern is only present with the lateral effect.  A simulated flat in Figure~\ref{fig:label13}, the lateral effect causes the flat variation, whereas the vertical effect does not.  Therefore, the lateral effect is responsible for astrometric shifts, flat variations, and PSF ellipticties, whereas the vertical effect leads to the variation of the PSF size.  Similar tree ring patterns were also studied in \cite{beamer2015}.

We can compare the magnitude of these two tree ring effects with published results.  We do not know intrinsically the dopant variation, $\alpha$, but note that the two lateral and vertical effects scale differently.  The magnitude of the lateral field effect is proportional to $\alpha t \frac{\frac{dE_{\perp}}{d \alpha}}{E_{\parallel}}$ where $t$ is the device thickness.  The parallel and perpendicular field is complex and depend on the doping level, bias voltage, and thickness.
Conversely, the relative change in the diffusion size due to the vertical effect is proportional to $\alpha \frac{\frac{dE_{\parallel}}{d \alpha}}{E_{\parallel}}$.  In the simulations we used 50\% doping variation and 100 $\mu m$ thickness.  This resulted in astrometric shifts of 10 $\mu m$, relative diffusion variations of 10\%, and a flat variation of 12\%.   \cite{magnier2018} measured astrometric shifts of 0.5 $\mu m$ and flat variations of 0.4\%.  Note that despite those are both from the lateral effect, there is a similar relative factor when comparing with the simulations (20 and 30 times, respectively).  \cite{magnier2018} measured a smear amplitude (square of the second moment), $s$, of 10 $\mu m^2$ for a typical PSF size, $\sigma$, of 16 $\mu m$.  So the relative diffusion variation would be $\frac{\Delta \sigma}{\sigma} = \frac{1}{\sigma} \left( \sqrt{\sigma^2 + s} - \sigma \right) = 2\%$.  This is about a factor of 5 smaller than the simulation and significantly less than the factor of 25 for the lateral effect implying a stronger parallel field dependence.  The stronger field dependence in this devices deserves further study.  Similarly, \cite{plazas2014} and \cite{bernstein2017} with DECam measure a flat variation of 3\% and a 3 $\mu m$ astrometric shift.  These results have a similar consistent relative factor of ~3 compared to the simulations.  The diffusion variation was not observed, but this should be more difficult to observe since DECam is much thicker at 250 $\mu m$.

We also can evaluate the wavelength-dependence of these effects.  In \cite{plazas2014} and \cite{bernstein2017}, the residual astrometric, photometric, and flat amplitude all decrease at long wavelength in observations with DECam.  This is because the photons convert at greater depth, and the effects of the lateral fields and diffusion are then decreased simply because they have less time to drift.  We simulated monochromatic flats at a range of wavelengths (350 nm, 450 nm, 550 nm, 650 nm, 750 nm, 850 nm and 950 nm) for a sensor of 250 $\mu m$.  The relative amplitude of the tree ring pattern relative to 350 nm was 1.0, 0.999, 0.997, 0.998, 0.930, and 0.663.  This is qualitatively consistent with the measurements of \cite{plazas2014} that had no effect for the r and i bands, and a ratio of appoximately 0.6 for the z band and 0.4 for the Y band.  The wavelength-dependence is smaller for thinner sensors, and it is also affected by other parameters than the thickness to a lesser degree.

\subsection{Brighter Fatter Effect and Signal vs. Variance}

\begin{figure*}[htb]
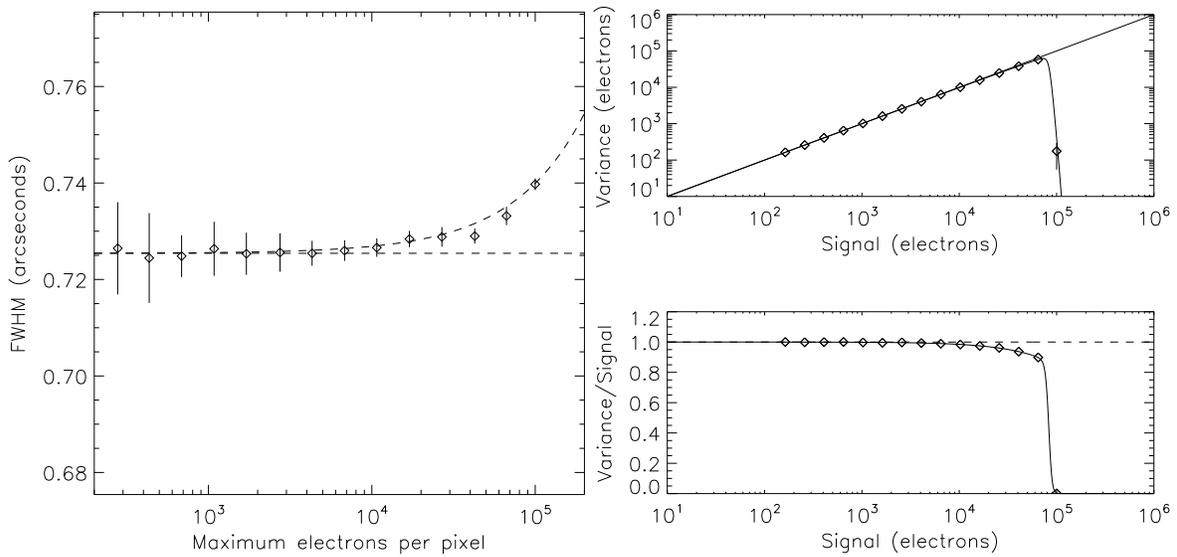

\begin{center}
\includegraphics[width=0.89\columnwidth]{f21.eps}
\includegraphics[width=0.89\columnwidth]{f22.eps}
\end{center}
\caption{\label{fig:label14} The observational effects of the distortion of the electric field due to accumulated charges on subsequent electrons.  The left plot shows the increase in PSF size as a function of the maximum electrons in a given pixel.  This demonstrates the intensity-dependent PSF behavior (brighter-fatter effect).  The right plot shows the variance of a simulated flat at different intensity values.  At low intensities the variance is equal to the intensity level as expected for uncorrelated Poisson noise, but at high intensities the electrons get preferentially shifted way from pixels with higher charge resulting in a sub-linear variance.}
\end{figure*}

The effect of the accumulated charge on the electric field and its influence over the lateral motion of subsequent electrons can be seen in two disperate effects:  1) the so-called brighter-fatter effect (\citealt{antilogus2014}; \citealt{gruen2015}; \citealt{rasmussen2015}) and 2) the non-linear relationship between signal and variance in CCDs (\citealt{downing2006}, \citealt{astier2019}).  The brighter-fatter effect refers to the observed sizes of objects being larger the brighter the source.  This happens because the distorted electric field due to the accumulated charge in the pixels in the core of a particularly bright source, tend to push the subsequent electrons further away.  This results in a larger PSF than the source would have if there were not as many photons.  This can be seen in Figure~\ref{fig:label14}, where we simulated a series of sources having different brightnesses, and plotted the measured PSF size.

The sub-linear relationship between the signal (total number of electrons) and its variance has been noted by \cite{downing2006}.  For pure Poisson noise, the variance should be linearly related to the signal.  However, it deviates from that relationship as the signal increases.  The distortion of the electric field due to the accumulated charges can cause this since the pixels that have a larger number of accumulated electrons due to positive statistical fluctuations will tend to induce larger lateral fields to keep new electrons away from those pixels.  This anticorrelation then tends to decrease the overall variance in the image and effectively smooths out the pattern.  Furthermore, this effect is intensity-dependent since at low fluxes the distorted electric fields from accumulated electrons is negligible and then the signal should be equal to the variance.  Figure~\ref{fig:label14} illustrates this by showing the sub-linear behaviour of the variance as a function of signal.  We performed a series of flat simulated at difference intensities to measure this effect.  The variance then drops to small value as it approaches the full well depth.  In this way, these two effects are simultaneously predicted from the physics of accumulated charge field distortion.  Thus, an intensity-dependent PSF size change of 2\% near saturation is directly related to a 10\% sub-linear variance for 100 $\mu m$ sensor and is consistent with measurements (\citealt{antilogus2014}).

\section{Conclusion and Future Work}

We demonstrated that complex non-ideal distortions in sensors can be implemented in an efficient Monte Carlo framework.  The simulation of photons incident on a sensor followed by the conversion of electrons and their response to electric fields can incorporate many of the commonly observed sensor distortion patterns.  Qualtitative results with fringing, field-free regions, lithography errors, and edge distortion were demonstrated.  Tree ring doping variation leads to two related effects.  The lateral field distortion which causes differential astrometric errors, PSF ellipticity errors, and flat varaiation, whereas the vertical field distortion causes PSF size errors.  We also demonstrated that intensity-dependent PSFs and sub-linear flat variance are both related to accumulated charges distorting the electric field pattern for subsequent electrons.

Future work will help to establish the accuracy of these methods as well as determine the variation of sensor distortions between different devices.  There are a number of additional less well-established effects resulting in non-uniformity that may be present in sensors that can be incorporated into this framework (see e.g. \citealt{antilogus2019}).  We also have limited this work to sensor effects that affect electrons before the electronic readout system.  PhoSim includes the most important aspects of these (\citealt{peterson2015}), but other details should be included in the future.

\acknowledgments
We thank the anonymous referee for helpful comments.  JRP acknowledges support from Purdue University and the Department of Energy (DE-SC00099223).  We thank Pierre Astier for helpful conversations.

\end{document}